\documentclass[twocolumn]{aastex63}
\usepackage{natbib, color, amsmath}
\graphicspath{{./figures}}
\begin{document}

\title{Mixed Source Region Signatures Inside Magnetic Switchback Patches Inferred by Heavy Ion Diagnostics}

\newcommand{\CfA}{\affiliation{Center for Astrophysics $\vert$ Harvard \& Smithsonian, 60 Garden Street, Cambridge, MA 02138, USA}}
\newcommand{\UMich}{\affiliation{University of Michigan
Department of Climate \& Space Sciences \& Engineering 2455 Hayward St.
Ann Arbor, MI 48109-2143, USA}}
\newcommand{\SwRI}{\affiliation{Space Science, Southwest Research Institute, 6220 Culebra Rd., San Antonio, TX, 78238, USA}}
\newcommand{\UCB}{\affiliation{Department of Physics, University of California, Berkeley, Berkeley, CA 94720-7300, USA}}
\newcommand{\SSL}{\affiliation{Space Sciences Laboratory, University of California, Berkeley, CA 94720-7450, USA}}
\newcommand{\MSSL}{\affiliation{Mullard Space Science Laboratory, University College London, Holmbury St. Mary, Dorking, Surrey, RH5 6NT, UK}}
\newcommand{\Iowa}
{\affiliation{Department of Physics and Astronomy, University of Iowa, Iowa City, IA 52242, USA}}
\newcommand{\Boulder}{\affiliation{Space Science Institute, Boulder, Colorado, USA}}

\author[0000-0002-8748-2123]{Yeimy J. Rivera}
\CfA

\author[0000-0002-6145-436X]{Samuel T. Badman}
\CfA

\author[0000-0002-7728-0085]{Michael L. Stevens}
\CfA

\author[0000-0001-5956-9523]{Jim M. Raines}
\UMich

\author[0000-0002-5982-4667]{Christopher J. Owen}
\MSSL

\author[0000-0002-5699-090X]{Kristoff Paulson}
\CfA

\author[0000-0001-6692-9187]{Tatiana Niembro}
\CfA

\author[0000-0002-4149-7311]{Stefano~A.~Livi}
\SwRI
\UMich

\author[0000-0003-1611-227X]{Susan T. Lepri}
\UMich

\author[0000-0002-9325-9884]{Enrico Landi}
\UMich

\author[0000-0001-5258-6128]{Jasper S. Halekas}
\Iowa

\author[0000-0002-8475-8606]{Tamar Ervin}
\UCB
\SSL

\author[0000-0003-4437-0698]{Ryan M. Dewey}
\UMich

\author[0000-0002-2576-0992]{Jesse T. Coburn}
\MSSL
\Boulder

\author[0000-0002-1989-3596]{Stuart D. Bale}
\UCB
\SSL

\author[0000-0001-6673-3432]{B.~L.~Alterman}
\SwRI

\begin{abstract}
Since Parker Solar Probe's (Parker's) first perihelion pass at the Sun, large amplitude Alfv\'en waves grouped in patches have been observed near the Sun throughout the mission. Several formation processes for these magnetic switchback patches have been suggested with no definitive consensus. To provide insight to their formation, we examine the heavy ion properties of several adjacent magnetic switchback patches around Parker's 11th perihelion pass capitalizing on a spacecraft lineup with Solar Orbiter where each samples the same solar wind streams over a large range of longitudes. Heavy ion properties (Fe/O, C$^{6+}$/C$^{5+}$, O$^{7+}$/O$^{6+}$) related to the wind's coronal origin, measured with Solar Orbiter can be linked to switchback patch structures identified near the Sun with Parker. We find that switchback patches do not contain distinctive ion and elemental compositional signatures different than the surrounding non-switchback solar wind. Both the patches and ambient wind exhibit a range of fast and slow wind qualities, indicating coronal sources with open and closed field lines in close proximity. These observations and modeling indicate switchback patches form in coronal hole boundary wind and with a range of source region magnetic and thermal properties. Furthermore, the heavy ion signatures suggest interchange reconnection and/or shear driven processes may play a role in their creation.

\end{abstract}


\section{Introduction} \label{intro}
Alfv\'en waves permeate interplanetary space in the form of correlated magnetic and velocity fluctuations in the solar wind \citep{Belcher1971}. The launch of Parker Solar Probe (Parker, \citealt{Fox2016}) enabled measurements of Alfv\'enic fluctuations to be taken closer to the corona than ever before, revealing that they assemble into well-defined large-scale structures in the near Sun environment. The collection of large amplitude Alfv\'enic fluctuations observed during Parker's perihelion passes are referred to as magnetic switchback \citep{Bale2019, Kasper2019, Phan2020}. Switchbacks are identified as large reversals/deflections in the magnetic field with Alfv\'enically-correlated spikes in velocity, maintaining a near-constant magnetic field magnitude and steady electron strahl pitch angle distribution direction. While the switchbacks are transient, lasting a few seconds to minutes, they are grouped into "magnetic switchback patches" that span a longitude of $\sim 5^{\circ}$ with similar spatial scales to supergranular cells at the photosphere \citep{Fargette2021}.    

Switchbacks were identified during Parker's first perihelion pass (at its closest approach of 37.7$R_{\sun}$, or solar radii) and have been pervasive in subsequent near-Sun encounters. However, studies find that their radial evolution destroys quintessential features used to identify them near the Sun \citep{Tenerani2021, Suen2023}. Therefore, they become less and less coherent with increasing heliocentric distance. 

Drivers of switchback formation are still debated but it is suggested that they are generated in the corona or in situ in the heliosphere, through a variety of mechanisms, such as interchange magnetic reconnection \citep{2004JGRA..109.3104Y, Fisk2020, 2020ApJ...903....1Z}, the expansion of the solar wind\citep{2015ApJ...802...11M, 2018ApJ...867L..26T, Squire2020, 2021ApJ...915...52S}, coronal jets \citep{2020ApJ...896L..18S} and velocity shear and footpoint motion \citep{2006GeoRL..3314101L, 2020ApJ...902...94R, 2020ApJ...888...68S, 2021ApJ...909...95S}.  Their compositional properties play a major role to track them back to their source region \citep{2007ApJ...660..901K, 2023ApJ...952...33H}. For instance, \cite{Bale2021} noticed increases of the alpha to proton abundance within patches of switchbacks interpreted as a consequence of magnetic reconnection processes in contrast to \cite{2022ApJ...933...43M} which found no consistent composition signatures after analyzing around a hundred switchback events during the 3rd and 4th Parker's encounters. 

Heavy ion measurements connected to switchbacks can provide diagnostics of coronal source region properties that can be used to disambiguate their source, a linkage which has not previously been made \citep{Rivera2022}. While Parker does not measure heavy ions ($Z>2$, where $Z$ is the atomic number or number of protons) for this type of analysis, Solar Orbiter \citep{Muller2020} carries a heavy ion resolving mass spectrometer, the Heavy Ion Sensor (HIS, \citealt{Livi2023}), as part of the Solar Wind Analyser (SWA, \citealt{Owen2020}) suite that can sample ion and elemental composition, albeit at larger heliocentric distances (closest distance at $\sim$60\,R$_{\sun}$) where switchback signatures are more ambiguous. However, this ambiguity can be mitigated through well-aligned spacecraft configurations, where observations of solar wind containing switchbacks taken by Parker closer to the Sun can be later sampled by Solar Orbiter to examine their heavy ion composition. 

The ionization states of heavy ions measured in the solar wind preserve signatures of the electron temperature and density of the corona they traveled through, making them excellent probes of the solar wind's source region characteristics. Ion abundances measured in situ throughout the heliosphere are governed by ionization and recombination processes happening in the corona and during the initial stages of outflow \citep{Landi2012, Boe2018, Gilly2020}. Ionization and recombination of ions in the plasma is determined by the plasma's electron temperature and density, and bulk flow. As collisions in the plasma are reduced due to expansion with increasing distance from the Sun, the ionization and recombination processes become less and less effective. Eventually ions convected in the solar wind reach their freeze-in distance, or heliocentric location where their relative abundances no longer change. In the solar wind, carbon and oxygen ions freeze-in closer to the Sun (1--1.5\,R$_{\sun}$) while iron ions can continue to ionize and recombine up to 3--5\,R$_{\sun}$ \citep{Landi2012}. Therefore, while other solar wind properties can change at different radial distances, heavy ion relative abundances are nearly static once the solar wind escapes the low and middle corona (\citealt{West2023} and references therein).

Elemental abundances in the solar wind are modulated by their source region's magnetic topology at the Sun. Decades of observations of elemental abundances derived from remote and in situ observations indicate strong first ionization potential (FIP) related fractionation in closed field regions of the Sun and associated slow speed wind \citep{Pottasch1963, Geiss1995, Raymond1997, vonsteiger2000, Feldman2000}. The FIP effect is the observed enhancement relative to photospheric composition of elemental abundances for atoms with an ionization energy below $\sim10$\,eV while elements above this threshold are either not affected or even depleted. The leading model that explains the FIP effect is connected to Alfv\'en waves reflecting/refracting across the chromoshperic-to-coronal boundary giving rise to the ponderomotive force. Depending on where Alfv\'en waves are reflected, the ponderomotive force will then act to preferentially transport charged low to mid-FIP elements (e.g. sulfur and carbon that sit at the low and high FIP energy threshold) into the corona \citep{Laming2004, Laming2012, Laming2019}. Once the solar wind is formed, elemental abundances remain fixed in the outflowing solar wind stream, maintaining the properties of its coronal source. The elemental abundances connected through remote and in situ observations have been important for identifying the sources of slow solar wind \citep{Brooks2015} and connecting transient structures from Sun to heliosphere \citep{Parenti2021}. 

As indicated above, we can connect heavy ion properties measured by Solar Orbiter to Parker during periods where they each measure the same solar wind streams. As such, we use a spacecraft conjunction between Parker's 11th perihelion pass (0.06\,au) and Solar Orbiter near the Sun-Earth line at 0.6\,au to connect heavy ion observations taken at Solar Orbiter to several subsequent switchback patches identified closer to the Sun with Parker. The Parker--Solar Orbiter conjunction has been examined in several cases \citep{Ervin2024, Rivera2024}. It is a special alignment, as both spacecraft cover similar Carrington longitudes and latitudes, which is rare. \cite{Ervin2024} examine the heavy ion composition at Solar Orbiter to map to distinctive solar sources of several fast and slow Alfv\'enic wind streams observed across an alignment in longitude spanning $\sim150^{\circ}$. The work indicated that the fast wind and Alfv\'enic slow wind in this encounter originated from coronal holes and their boundaries, respectively. \cite{Rivera2024} examines the radial evolution of a switchback patch during the alignment, included in this study, finding their contribution to be significant and necessary to the full description of the heating and acceleration experienced by the fast solar wind. \cite{Rivera2024} analyzed the elemental and ion composition of the switchback patch, finding its Fe/O abundance ratio to be in line with fast solar wind elemental composition, albeit indicating some low-FIP enhancement. 

In this work, we expand on the analysis from \citet{Rivera2024}, to examine the ion and elemental composition of adjacent switchback patches observed with Parker and eventually measured at Solar Orbiter. Through ballistic backmapping methods used previously \citep{Ervin2024, Rivera2024}, we align switchback patch observations at Parker to O$^{7+}$/O$^{6+}$, C$^{6+}$/C$^{5+}$, and Fe/O measurements connected to the same mapped source surface longitude. We find that heavy ion measurements show variation across neighboring patches in both ion and elemental composition, indicating changes associated with source region conditions and locations at the Sun. The Fe/O variability and electron temperature connected to the patches spans between typical coronal hole and quiet Sun properties (in contrast to active regions at the Sun which exhibit a much more extreme FIP bias), indicating a source with a mixture of open/closed magnetic field topology such as coronal holes boundaries.

The paper is organized as follows: Section \ref{sec:streammatching} describes the observations from Parker and Solar Orbiter and the ballistic mapping technique. Section \ref{sec:Results} presents the heavy ion measurements characteristics within the switchback patches observed at Parker. Section \ref{sec:Discussion_conclusions} summarizes the results.

\section{Stream Matching} \label{sec:streammatching}

\subsection{Parker Solar Probe and Solar Orbiter observations}
At Parker, we examine observations from the SPAN-Ai instrument to compute the bulk solar wind velocities \citep{Livi2022}. SPAN-Ai couples an electrostatic analyzer and time-of-flight (TOF) component to resolve incident angle, mass-per-charge, and energy-per-charge of incoming ions. The mass discrimination made possible by the TOF analyzer allows for the identification of separate ions species in the solar wind, mainly the most abundant species -- proton and alpha particles, or He$^{2+}$. We include electron measurements from SPAN-e \citep{Whittlesey2020} where the electron temperature is derived by methods described in \cite{Halekas2020}. Additionally, the 3D magnetic field components are measured by the fluxgate magnetometer (MAG) on FIELDS at 4\,vectors/cyc that captures the rapid changes in the magnetic field \citep{Bale2016}. 

At Solar Orbiter, measurements of the magnetic field are taken by another fluxgate magnetometer (MAG) at 8\,vectors/s \citep{Horbury2020}. Observations of protons, alpha particles, and heavier ions across this period were taken by Proton-Alpha System (PAS) and HIS that are part of the SWA suite \citep{Owen2020}. The proton velocities were measured with the PAS instrument with 4\,s full scan 3D particle distributions.  Observations of the heavier ions ($Z>2$) were measured at a 30\,s cadence and accumulated to 10 minute resolution using the TOF mass spectrometer, HIS \citep{Livi2023}.

\subsection{Ballistic Mapping}

As in \citet{Ervin2024, Rivera2024}, Solar Orbiter and Parker measurements are cast from time series to spatial coordinates termed \lq{}source surface longitude\rq{} by using ballistic mapping \cite{Nolte1973,Stansby2019,Badman2020} to associate each measurement with a heliographic location at the edge of the corona, at $2.5R_{\sun}$. This process models plasma parcel streamlines as archimedean spirals in the solar corotating frame using the measured solar wind velocity at each spacecraft and assuming the parcel moves at a constant radial speed. In comparison to models which account for coronal corotation and acceleration, this approximation has recently been quantified to yield results with errors under $5^\circ$ when starting from 1\,au, and monotonically decreasing with close approach to the Sun \citep{Dakeyo2024}. Here, our measurements are taken at 0.6\,au (Solar Orbiter) and 0.063\,au (Parker) and so are expected to be even smaller. We study statistically a region of source surface longitude of 70$^\circ$ in width and identify switchback patches of width similar or greater than $5^\circ$.

\section{Heavy Ion Observations} \label{sec:Results}
\begin{figure*}[]
	\centering
	\includegraphics[width=\linewidth]{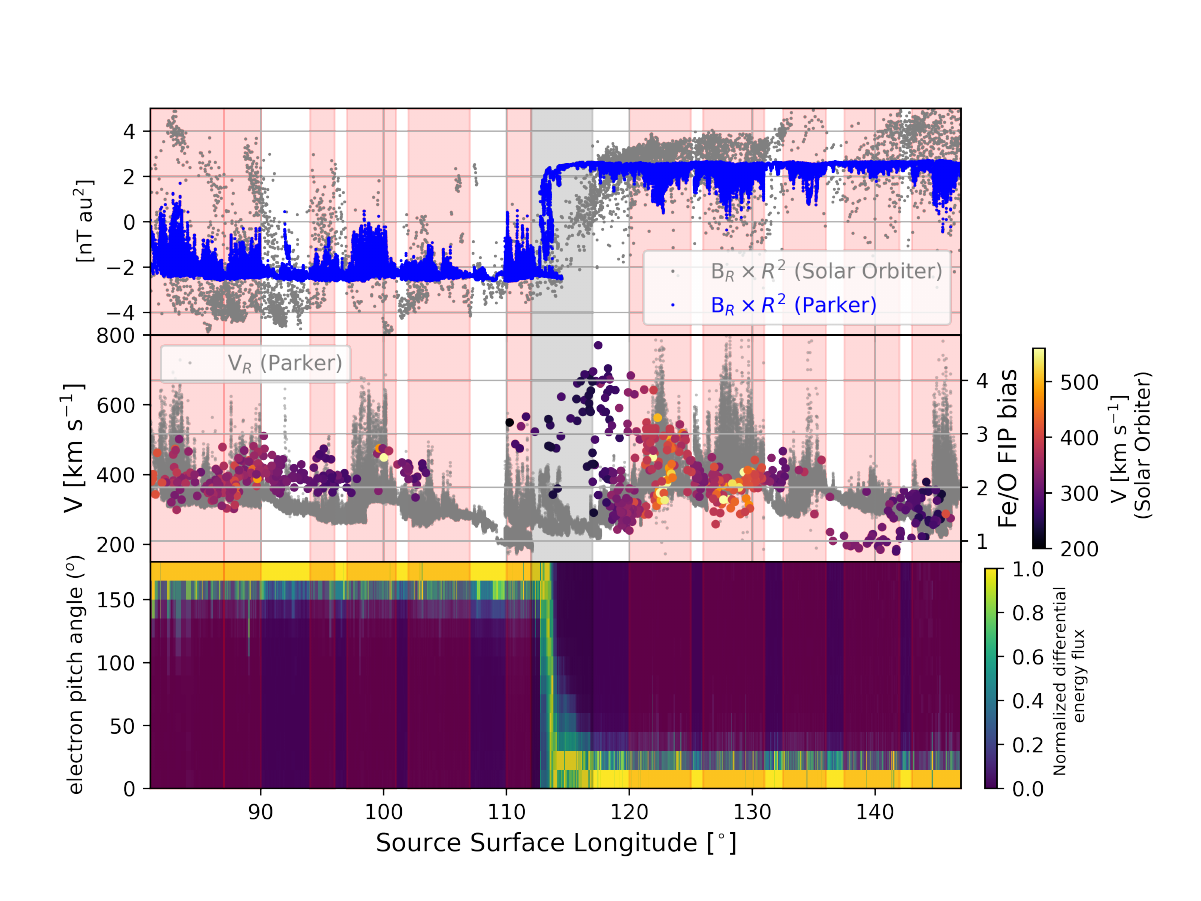}
	\caption{Multi-panel plot showing the switchback properties at Parker and compositional properties at Solar Orbiter across source surface longitude. Top: radial magnetic field, B$_R$ multiplied by heliocentric distance squared for Parker (blue) and Solar Orbiter (gray). Middle: left vertical axis shows the radial speed at Parker (gray) and the right axis shows the Fe/O ratio normalized to the Sun's photospheric abundance, as FIP bias. The FIP bias is colored by the wind speed at Solar Orbiter. Bottom: Normalized differential energy flux of 432.69\,eV electrons measured at Parker. Shaded red regions identify individual switchback patches characterized by the speed, magnetic field, and electron e-PAD signatures. The shaded gray region is the current sheet crossing.}
	\label{fig:elemental_comp}
\end{figure*}

\subsection{Elemental Composition}
Figure \ref{fig:elemental_comp} shows a three panel plot of the properties of Parker and Solar Orbiter ballistically backmapped across source surface longitude. The top panel shows the magnetic field radial component multiplied by the radial distance squared at each spacecraft for Solar Orbiter (gray) and Parker (blue) in order to account for the radial dependence in the radial component of the magnetic field between the two spacecraft. The middle panel shows the bulk speed at Parker (gray) along with the Fe/O FIP bias colored by the bulk speed at Solar Orbiter. The Fe/O FIP bias is computed by normalizing the relative elemental abundances of Fe and O to their photospheric elemental composition from \cite{Asplund2021}. A FIP bias above 1 indicates some enhancement of Fe or depletion of O in the plasma and is interpreted as indicating the plasma has spent time on closed magnetic loops prior to escaping. The bottom panel shows the column normalized electron pitch-angle distribution (e-PAD) of suprathermal electrons across $0-180^{\circ}$ in the energy range 432.69eV. The red shaded regions denote the individual, coherent switchback patches across source surface longitude $81-147^{\circ}$ spanning the full day of 2022\,February\,25 at Parker and spanning 2022\,February\,23 to March\,4 at Solar Orbiter. The switchback patches were identified through their typical characteristics at Parker \citep{Bale2021,Fargette2021}: 
\begin{itemize}
  \item large fluctuations of radial magnetic field component correlated to fluctuations in speed
  \item constant magnetic field magnitude across the patch structure
  \item steady field-aligned electron beam
  \item consistent longitudinal width around $5^\circ$
\end{itemize} 

A current sheet crossing occurs midway through Parker's closest approach, showing a change in the magnetic field polarity inferred from the e-PAD, and a local speed minimum near $113^{\circ}$. The current sheet crossing is also observed at Solar Orbiter, as indicated by the change in the radial magnetic field components also near $113-117^{\circ}$. The well-aligned current sheet crossings at both spacecraft demonstrates an accurate mapping of flows between Parker (at $\sim$13.5\,R$_{\sun}$) and Solar Orbiter (at $\sim$130\,R$_{\sun}$) near the current sheet. Therefore, we focus on characterizing switchback patches embedded in solar wind within $\sim30^{\circ}$ of the current sheet crossing where we have the greatest confidence that Parker and Solar Orbiter sampled the same streams.  

As shown in the middle panel of Figure\,\ref{fig:elemental_comp}, we find that the Fe/O FIP bias is generally observed to be between 2--3 within the switchback patches spanning this longitude range, with the exception of the small amplitude patch near $137-142^{\circ}$ with a photospheric elemental composition, near 1. Generally, the variability of the Fe/O FIP bias measured across wind speed and the solar cycle spans, $1.7\pm1.1$ (slow) to $0.8\pm0.44$ (fast) in solar max and $1.4\pm1.1$ (slow) to $0.9\pm0.34$ (fast) in solar minimum conditions \citep{Lepri2013}. Therefore, the mean values observed are closer to what is expected for slow solar wind, for solar max or minimum. However, the large variability within the observed patches covers a mixture of photospheric to coronal abundances connected to both open and closed field structures at the Sun.

\subsection{Electron Temperature and Ion Abundances} 
\begin{figure*}[]
	\centering
	\includegraphics[width=\linewidth]{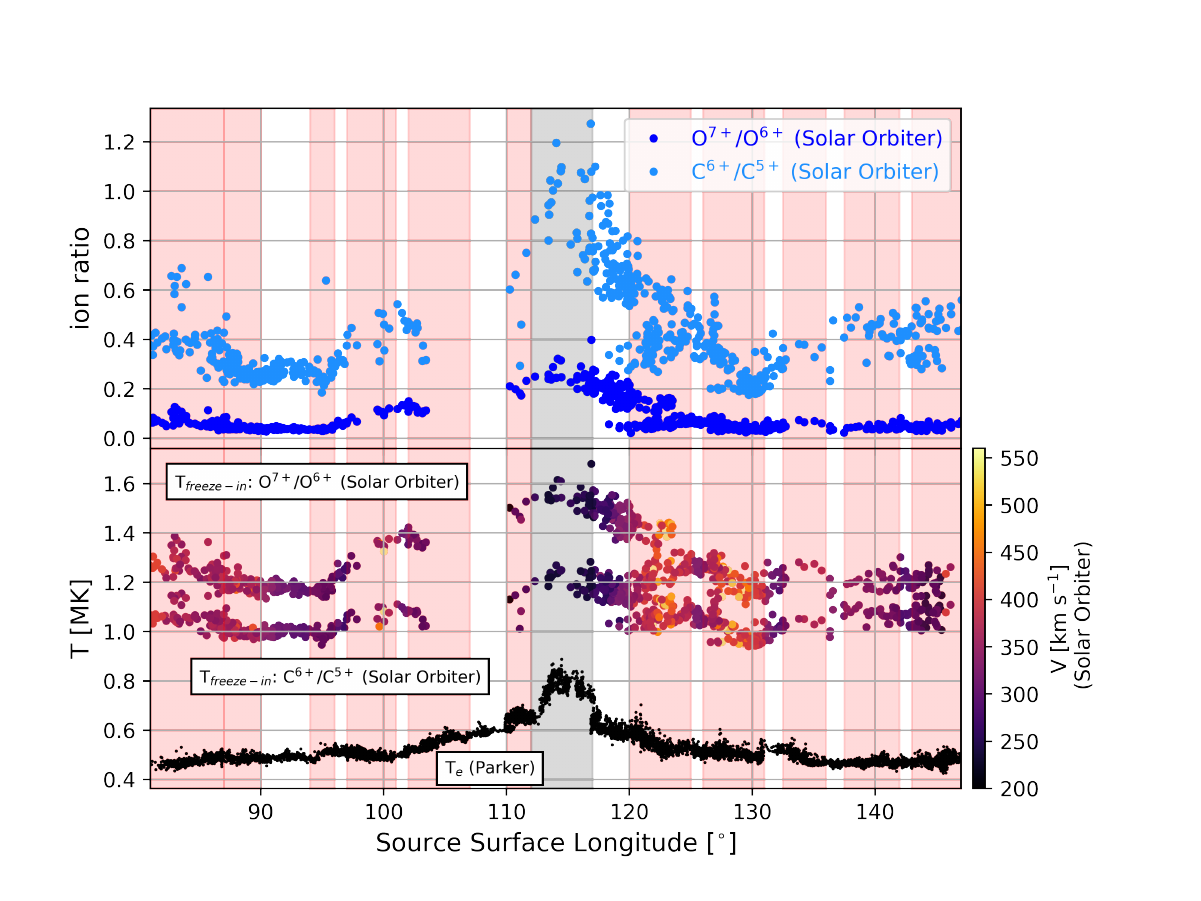}
	\caption{Two panel plot showing the ion ratio and associated electron temperature properties of the solar wind. Top: relative abundance ratio of O$^{7+}$/O$^{6+}$ and C$^{6+}$/C$^{5+}$ measured at Solar Orbiter, bottom: derived electron temperature from O$^{7+}$/O$^{6+}$ and C$^{6+}$/C$^{5+}$ formation temperatures from Solar Orbiter and electron temperature computed in situ at Parker. Shaded regions are the same as Figure\,\ref{fig:elemental_comp}}	
 \label{fig:ion_comp}
\end{figure*}

Figure\,\ref{fig:ion_comp} shows the ion ratios and the associated freeze-in temperature across the same source surface longitude as Figure\,\ref{fig:elemental_comp}. The top plot shows the ballistically backmapped relative abundance ion ratios of O$^{7+}$/O$^{6+}$ and C$^{6+}$/C$^{5+}$ at Solar Orbiter. The bottom plot shows the freeze-in temperature, T$_{freeze-in}$, derived from the ion formation temperatures, colored by the bulk speed at Solar Orbiter, under the assumption of ionization equilibrium conditions, as well as the computed electron temperature at Parker (black). The shaded regions are the same switchback patches identified from properties in Figure\,\ref{fig:elemental_comp} and the shaded black region shows the current sheet crossing location. The T$_{freeze-in}$ is derived by computing the ratio of the relative abundances of each ion as a function of electron temperature using CHIANTI v.10, detailed in \cite{delzanna2021}. T$_{freeze-in}$ is the temperature where the computed curve of the ion ratio matches the ion ratio measured in situ. The accuracy of the derived freeze-in temperature from each ion ratio relative to a meaningful coronal temperatures relies on the ions freezing-in while in ionization equilibrium. Therefore we do not claim in each case that the temperature is definitively reflective of the source region at a specific place in the corona.

Generally, the ion ratios in the top panel of Figure\,\ref{fig:ion_comp} show profiles that increase, nearly symmetrically, as they approach the current sheet (near $115^{\circ}$), with their highest values within the current sheet as typically observed \citep{Rivera2021, Lynch2023}. The associated freeze-in temperature derived from both ratios show correlated behavior and a similar profile to the electron temperature measured independently in situ at Parker (black). We compute a correlation coefficient (ratio of the covariance of the two timeseries divided by the standard deviation of each timeseries multiplied together) between the ion parameters and the electron temperature at Parker to quantify their correlated behavior. We downsample each series into 2 degree averaged longitudinal bins for comparison. We find the correlation coefficient of electron temperature at Parker and T$_{freeze-in}$ from O$^{7+}$/O$^{6+}$ to be 0.87, while the electron temperature at Parker and the O$^{7+}$/O$^{6+}$ ion ratio shows an even higher correlated value of 0.93. Similarly for the carbon ratios, we find the correlation coefficient of the electron temperature at Parker and T$_{freeze-in}$ from C$^{6+}$/C$^{5+}$ is 0.80, while the electron temperature at Parker and the C$^{6+}$/C$^{5+}$ ion ratio is 0.87. We also compute the correlation coefficient of O$^{7+}$/O$^{6+}$ and C$^{6+}$/C$^{5+}$ that returns a similar correlation coefficient of 0.91 showing a strong correlated behavior overall.

The freeze-in temperature derived from oxygen charge states shows a systematic 0.2\,MK difference across the entire range relative to that derived from carbon. The changes in the freeze-in temperature would be consistent with a difference in freeze-in radial distances between the carbon and oxygen ratios \citep{Landi2012}. However, some deviation from ionization equilibrium cannot be ruled out as responsible for the shift. 

Overall, the ion ratios are less structured compared to the Fe/O FIP bias, and show small variation in both the freeze-in and local electron temperature across the switchback patches. The electron temperature measured at Parker and the ion ratios/freeze-in temperatures show similar substructure. In particular, we find correlated changes in the electron temperature profile at Parker and ion ratios across several switchback boundaries. For instance, all three profiles show a steady profile between source surface longitude 87--94$^{\circ}$ with a steady rise in temperature after that point. We also observe a concurrent increase in temperature near $131^{\circ}$, likely associated with a change in source region properties across solar wind streams. All profiles indicate a steady electron temperature across two switchback patches spanning 137--147$^{\circ}$.

\begin{figure*}[]
	\centering
	\includegraphics[width=\linewidth]{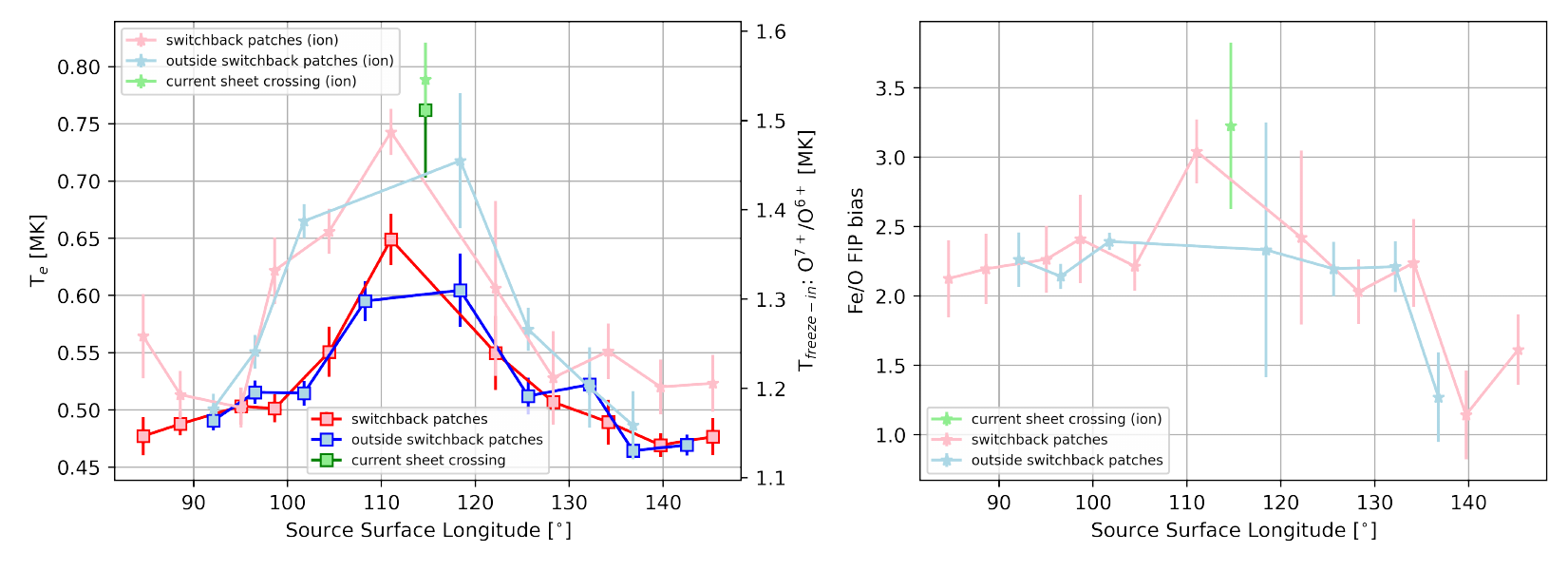}
	\caption{The left plot shows the mean electron temperature computed within (red) and outside (blue) switchback patches and at the current sheet (green) with the standard deviation across the longitude range. The electron temperature computed at Parker is shown in the solid lines associated with the left vertical axis, and the ion ratio derived freeze-in temperature in the light colors associated with the right vertical axis. The plot on the right shows the mean and standard deviation of the Fe/O FIP bias for the different cases.}	
 \label{fig:ion_temp}
\end{figure*}

We find larger variation in electron temperatures across longitude patches than within individual patches. The left plot of Figure\,\ref{fig:ion_temp} shows the mean in situ electron temperature from Parker across individual switchback patches (red), in between the switchback periods (blue), and spanning the current sheet (green) with the lighter colors showing the same classification except for the freeze-in temperature derived from O$^{7+}$/O$^{6+}$. The vertical bars in the plot are the standard deviation of the temperature within the different longitude patches and indicate the variability of the temperature within that region. The figure shows that at Parker, the typical standard deviation within individual patches (a single red box) is much smaller (on average $\sim$0.018\,MK) compared to the variation in electron temperature of all the patches across longitude (variation across all red boxes in Figure \ref{fig:ion_temp}), with a standard deviation $\sim$0.05\,MK.  The same is observed in the freeze-in temperature. We note that the period covered by the Parker measurements spans a radial distance of $2.6R_{\Sun}$ between $13.28-15.9R_{\Sun}$ which is too small \citep{Dakeyo2022} to explain the variation seen in in situ electron temperature in Figure \ref{fig:ion_temp}. Therefore, we attribute the variation in in situ electron temperature at Parker to changes in the source region conditions rather than radial evolution.

We find Fe/O FIP bias shows large variability across different patches, as shown in the pink curve of the right plot in Figure\,\ref{fig:ion_temp}. The vertical bars are computed in the same manner as the left plot. However, the FIP bias is similar to outside of the switchback patches suggesting their source is similar to the surrounding non-switchback solar wind.

\section{Summary and Conclusions} \label{sec:Discussion_conclusions}
We examine the heavy ion composition of several adjacent switchback patches measured during Parker's 11th perihelion pass centered on 2022\,February\,25. Encounter 11 occurred during a well-aligned conjunction with Solar Orbiter that sampled a similar longitude and latitude as Parker. The spacecraft conjunction enabled young solar wind measured in situ by Parker (at 13.5\,R$_{\sun}$) to be connected to solar wind streams measured at Solar Orbiter (at 130\,R$_{\sun}$). Through ballistic backmapping from both spacecraft to the source surface at 2.5\,R$_{\Sun}$, the properties at Parker were aligned in longitude to measurements at Solar Orbiter. We used a current sheet crossing as indication of backmapping accuracy that served as an anchor point to compare the surrounding solar wind streams. We mapped switchback structures observed at Parker to measurements of heavy ions observed at Solar Orbiter to determine the Fe/O abundance and electron temperature properties within each patch. Given Parker's proximity to the Sun, switchback patches appear more coherent, making them more easily identifiable than at larger distances where they are less well organized, however Parker does not measure the needed heavy ion signatures. Therefore, the alignment and use of both spacecraft was crucial to mapping their heavy ion properties. 

We find the elemental composition (Fe/O) and the ion ratios suggest some variability in the solar source of the neighboring switchback patches. The Fe/O FIP bias across the adjacent patches show a range between 1--3 suggesting their source is undergoing the FIP effect to varying degrees. However, the strong variability between and within each switchback patch suggests distinct sources, or some combination of coronal regions, are related to the individual patches. 

Similarly, but to a lesser extent, the ion ratios indicate some sub-structure across adjacent patches. The derived freeze-in temperatures from the carbon and oxygen ratios maintains a similar profile to the ion ratios themselves as well as to the in situ electron temperature measured by Parker. The strong correlation in features observed between ion ratios and the ion ratio derived coronal electron temperature from Solar Orbiter observations and in situ electron temperature measured independently at Parker indicate: 1) a compelling connection between solar wind observations at Parker and Solar Orbiter, 2) changing source conditions across adjacent patches, and 3) cooler electron temperature (in situ) compared to current sheet plasma.

We conclude that heavy ion measurements show notable variation across neighboring patches in both ion and elemental composition. The elemental composition indicates the Fe/O FIP bias spans between typical coronal hole and quiet Sun composition, suggesting both as their source \citep{DelZanna2019} in contrast to active regions that contain the most extreme low-FIP enhancements (FIP bias $>4$) and the hottest coronal plasma, $3-10$MK \citep{Yoshida1996, Feldman2000, Widing2001, Reale2010}. We note we do not see examples of switchback patches forming in wind with composition properties of current sheet plasma, active region outflow, or pristine fast wind. The range of Fe/O FIP bias measured within and across switchback patches indicates their solar source is a combination of open (not low-FIP enhanced) and closed (low-FIP enhanced) field structures. This result, along with modeling in \cite{Ervin2024, Rivera2024} throughout this timeframe, connects the switchback patches to open/closed field boundaries at the edge of coronal hole where a transition of photospheric to low-FIP enhanced Fe/O values are observed \citep{Stakhiv2015, Xu2015}. Extended equatorial coronal holes were a predominant feature on disk during this time interval in line with the switchback patches mapping to open/close field regions \citep{Ervin2024, Rivera2024}. The range of compositional values indicates a diversity of source thermal and magnetic conditions within this distinct source type, all of which enable the formation of switchback patches.

We note that because the main coronal feature associated with the switchback patches examined in this study were mapped to extended coronal holes, it is difficult to distinguish from this interval if switchback patches are simply reflecting the compositional characteristics of the background solar wind they are embedded in, or if they \textit{require} the observed source region properties seen in this study. Future work should identify similar aligned observations with more clear extremes in source region properties, such as those with large equatorial coronal holes, to verify if switchbacks patches in these cases diverge from background solar wind compositional signatures.

\subsection{Switchback Formation Processes}
As listed in the introduction, there are several switchback formation processes theorized. Here we compare the predicted observables from some of these processes to the heavy ion and electron temperature characteristics of switchback patches we find in this study, which were:,

\begin{enumerate}
  \item Open and closed field FIP bias characteristics
  \item Steady ion abundances congruent with non-switchback solar wind
  \item Steady electron temperature similar to non-switchback solar wind temperature
\end{enumerate} 

The solar wind bearing the switchback patches in this interval has been mapped to open/closed boundaries \citep{Ervin2024} that are ideal locations for interchange reconnection to occur, a phenomenon featured in many theories of switchback formation \citep{Fisk2020, Bale2023}. However, the lack of evidence for strongly enhanced electron temperatures from Parker (above surrounding solar wind conditions) suggests any strong heating occurring in such reconnection is either averaged out across the stream long before the plasma reaches Parker or happening above the ion's freeze-in height such that the ions do not reflect the heating \citep{Scott2022}. 

Similarly, velocity shear driven formation processes would be in line with a mixed FIP bias observation since it requires adjacent fast and slow solar wind to drive switchback formation \citep{2020ApJ...902...94R, 2020ApJ...888...68S, 2021ApJ...909...95S}. However, this formation process would also ultimately map to regions of open/closed field such as coronal hole boundaries, as in the interchange reconnection case, where slow and fast solar wind form in close proximity. For cases of similar source regions, composition measurements do not immediately distinguish these two processes.

Lastly, coronal jets would indicate the switchbacks form through the ejection of mini-filament flux ropes \citep{2020ApJ...896L..18S}. However, similar to larger scale flux ropes, this would be indicated through enhancements to the ion abundances and Fe/O FIP bias that would result in a higher mean value compared to background solar wind properties, however that is not observed in this interval.

\acknowledgments

Y.J.R., S.T.B., T.N. and K.P. were partially supported by the Parker Solar Probe project through the SAO/SWEAP subcontract 975569.
B.L.A. acknowledges support from NASA contract NNG10EK25C and NASA grants 80NSSC22K0645 (LWS/TM) and 80NSSC22K1011 (LWS).
E.L. acknowledges support from NASA grant 80NSSC20K0185.

{Parker Solar Probe was designed, built, and is now operated by the Johns Hopkins Applied Physics Laboratory as part of NASA’s Living with a Star (LWS) program (contract NNN06AA01C). Support from the LWS management and technical team has played a critical role in the success of the Parker Solar Probe mission.

Solar Orbiter is a mission of international cooperation between ESA and NASA, operated by ESA. Solar Orbiter SWA data were derived from scientific sensors that were designed and created and are operated under funding provided by numerous contracts from UKSA, STFC, the Italian Space Agency, CNES, the French National Centre for Scientific Research, the Czech contribution to the ESA PRODEX programme and NASA. SO SWA work at the UCL/Mullard Space Science Laboratory is currently funded by STFC (Grant Nos. ST/W001004/1 and ST/X/002152/1).  The SWA-HIS team acknowledges NASA contract NNG10EK25C.  
}

\end{document}